# Conical Multi-Beam Phased Arrays for Air Traffic Control Radars

P. Rocca,[(1)(2)] *Senior Member, IEEE*, N. Anselmi,[(1)] *Member, IEEE*, M. A. Hannan,[(1)] and A. Massa,[(1)(3)(4)] *Fellow, IEEE*

[(1)] *CNIT* - "University of Trento" ELEDIA Research Unit
Via Sommarive 9, 38123 Trento - Italy
E-mail: {*paolo.rocca*, *nicola.anselmi.1*, *mohammadabdul.hannan*, *andrea.massa*}@*unitn.it*
Website: *www.eledia.org/eledia-unitn*

[(2)] *ELEDIA Research Center* (*ELEDIA@XIDIAN* - Xidian University)
P.O. Box 191, No.2 South Tabai Road, 710071 Xi'an, Shaanxi Province - China
E-mail: *paolo.rocca@xidian.edu.cn*
Website: *www.eledia.org/eledia-xidian*

[(3)] *ELEDIA Research Center* (*ELEDIA@UESTC* - UESTC)
School of Electronic Engineering, Chengdu 611731 - China
E-mail: *andrea.massa@uestc.edu.cn*
Website: *www.eledia.org/eledia-uestc*

[(4)] *ELEDIA Research Center* (*ELEDIA@TSINGHUA* - Tsinghua University)
30 Shuangqing Rd, 100084 Haidian, Beijing - China
E-mail: *andrea.massa@tsinghua.edu.cn*
Website: *www.eledia.org/eledia-tsinghua*

# Conical Multi-Beam Phased Arrays for Air Traffic Control Radars

P. Rocca, N. Anselmi, M. A. Hannan, and A. Massa

**Abstract**

The design of conical phased array antennas for air traffic control (*ATC*) radar systems is addressed. The array architecture, which is controlled by a fully digital beam-forming (*DBF*) network, is composed by a set of equal vertical modules. Each module consists of a linear sparse array that generates on receive multiple instantaneous beams pointing along different directions in elevation. To reach the best trade-off between the antenna complexity (i.e., minimum number of array elements and/or radio frequency components) and radiation performance (i.e., matching a set of reference patterns), the synthesis problem is formulated in the Compressive Sampling (*CS*) framework. Then, the positions of the array elements and the complex excitations for generating each single beam are jointly determined through a customized version of the Bayesian *CS* (*BCS*) tool. Representative numerical results, concerned with ideal as well as real antenna models, are reported both to validate the proposed design strategy and to assess the effectiveness of the synthesized modular sparse array architecture also in comparison with conventional arrays with uniformly-spaced elements.

# 1  Introduction

Surveillance systems are crucial elements in airports for the correct execution of all terminal functions and procedures such as airplanes take off and landing, air traffic management, and terminal security. Nowadays, most air traffic control (*ATC*) radars use ground stations, equipped with large reflectors and/or array antennas installed on mechanical rotating pedestals, to yield a 360 [deg] view for tracking the targets all around. Such a technology, in place since the 70' [1], is currently close to upgrade and novel solutions, based on cylindrical or multi-face planar phased arrays (*PA*s) [2]-[5], are currently investigated. Thanks to the rapid scan, the clutter suppression, the multi-beam generation, and the multi-function capabilities (e.g., joint *ATC* and weather radar functionalities [6]-[8]), electronically scanning *PA*-based solutions seem to be promising candidates for the next generation of *ATC* radar systems. Indeed, future *ATC* radars are expected to assure safe and resilient performance thanks to a highly-reliable detection and tracking based on accurate 3D scanning, real-time adaptive nulling, pattern shaping, and self-calibration enabled by fully *DBF* arrays where the signal received by each radiating element is independently digitized with an analog-digital converter (*ADC*) [9][10]. Compared to an analog beam-forming (*ABF*), the *DBF* has the advantage of simultaneously generating multiple beams on receive with arbitrary shapes. Moreover, *DBF* allows a very accurate pointing over wide bandwidths by avoiding the beam squint effects thanks to the use of true time delays [11].

However, *ATC* radars require narrow beams to achieve high resolution and, consequently, they need *PA*s with large apertures and a huge number of radiating elements as well as *ADC*s. Although the cost, the size, and the power consumption of *ADC*s are continuously decreasing, the use of *PA*s with classical/fully-populated (*FP*) architectures (i.e., array layouts with uniformly-spaced radiating elements, each one equipped with a dedicated *ADC*) would result too expensive for commercial applications. Therefore, scalable and modular *DBF* implementations are of great interest for future fully-digital *ATC PA*-based radars. In this framework, innovative unconventional array architectures have been recently proposed to implement better cost-performance trade-off solutions [12]. For instance, clustered/sub-arrayed *PA*s have been considered for reducing the number of control points by gathering multiple elements over a single input-output port to minimize the production and maintenance costs [13]-[22]. The main issue when re-sorting to clustered arrays is the unavoidable presence of undesired secondary lobes (i.e., the *quantization lobes* [23][24]), whose amplitudes grow widening the bandwidth as well as the scan angle from the array broadside. To cope with this drawback, irregular arrangements of the array clusters, leading to an aperiodic distribution of the sub-array phase centers, have been exploited and the optimization of the sub-array configurations (i.e., sub-array position/size/shapes and excitations) has been carried out by means of various strategies based on evolutionary algorithms [14][16][17][19]-[21], integer programming [18], Compressive Sensing (*CS*) [15], and random techniques [22].

Alternatively, another effective approach to simplify the *PA* architecture is array sparsening [25]-[33]. By jointly optimizing the inter-element distances between the array elements and the corresponding excitation weights, the number of array elements is minimized, while fitting user-defined radiation performance. Deterministic [26][32] and stochastic techniques [25][27] as well as *CS*-based methods [28]-[31][33][34] have been published in the recent scientific literature to synthesize sparse arrays.

This paper presents a *DBF PA* antenna for *ATC* radar systems, which has been preliminary introduced in [35], that benefits from both the modularity and the sparseness of the array layout. More specifically, the array is composed by sparse vertical planks (i.e., sub-array modules), to minimize the number of antenna elements and of the corresponding *ADC*s, positioned side-by-side over a truncated conical surface. The choice of a conical shape is quite natural since the surface is already tilted towards the sky, thus a lower elevation scanning is needed [36][37], while the *DBF* simultaneously generates multiple beams to cover the required elevation range. Thanks to the modular architecture along the vertical direction, the array affords a 360 [deg] view along the azimuth plane. Moreover, the processing of the signals collected by a contiguous subset of the array planks guarantees uniform radar performance.

To determine the linear sparse arrangement of the elementary radiators of the vertical plank and the corresponding excitation sets, the synthesis problem is cast as an optimization one aimed at generating multiple beams, as close as possible to those radiated by an ideal/reference *FP* faithfully fulfilling the user requirements, while jointly minimizing the number of elements. Towards this end, a sparse-regularization technique based on the Bayesian *CS* (*BCS*) technique is here exploited because of its proved efficiency and effectiveness in finding the sparsest linear [28][29], planar [30] and conformal [31] array layouts that minimize the pattern matching misfit. However, none of previous *BCS* synthesis methods has previously dealt with the joint generation of multiple radiation patterns. Therefore, this work introduces a novel synthesis strategy based on a Multi-Task (*MT*) [29][30][31] implementation of the *BCS* where the condition of common positions for the array elements is enforced, while optimizing an independent set of complex excitations for each beam, by minimizing the *a-posteriori* probability for the array to match the target patterns by means of a fast Relevance Vector Machine (*RVM*) solver [38].

To the best of the authors knowledge, the main novelties of this work with respect to the state-of-the-art literature include (*i*) the introduction of an innovative modular architecture (i.e., a conformal/conical multi-beam sparse receiving array) for *ATC* radar systems; (*ii*) the theoretical formulation of the design problem within the *CS* framework to yield a solution that assures the best compromise/trade-off between the closeness/fulfilment of the user/operative radiation requirements and the reduction of the architectural complexity (i.e., the minimum number of elements/*ADC*s); (*iii*) the development of a synthesis tool based on a customized implementation of the *BCS* to jointly synthesize the array layout and the sets of complex excitations for generating the multiple beams.

The rest of the paper is organized as follows. The synthesis of conical fully-digital arrays for *ATC* radars is mathematically formulated in Sect. 2, where the proposed *MT-BCS* based solution method is described, as well. Section 3 presents a set of selected numerical results to validate the synthesis method and to assess the effectiveness of the arising layouts also considering realistic antenna models. Eventually, conclusions follow (Sect. 4).

## 2 Mathematical Formulation

Let us consider a conical phased array (*CPA*) of $P$ radiating elements located on the surface of a truncated cone with axis along the $z$ Cartesian coordinate and circular bases lying on the ($x$,$y$)-plane with radii $R$ (major base radius) and $r$ (minor base radius), respectively (Fig. 1). Such a *CPA* is composed by $N$ vertical planks of $M$ elements ($P \triangleq M \times N$) placed side-by-side at a constant distance $d_c$ measured along the circular perimeter of the minor base of the cone. To yield a modular *CPA* architecture, all $N$ planks are equal and they consist of a sparse linear distribution of the elementary radiators over a regular lattice, $d_m$ ($m = 1, ..., M - 1$) being the inter-element distance between the $m$-th and $(m + 1)$-th adjacent elements, while $l$ ($l \triangleq \sum_{m=1}^{M-1} d_m$) is the length of the plank [i.e., the distance between the first (i.e., $m = 1$) and the last (i.e., $m = M$) elements along the plank axis]. Thus, $r = \frac{d_c \times (N-1)}{2\pi}$ and $R = r + l \times \cos \theta_S$, $\theta_S$ being the cone slant angle (Figs. 1-2), while the coordinates of the position of the $m$-th ($m = 1, ..., M$) element in the $n$-th ($n = 1, ..., N$) array-column/plank turn out to be

$$\begin{aligned} x_{m,n} &= R \times \cos(n\psi_c) - l_m \cos\theta_S \\ y_{m,n} &= R \times \sin(n\psi_c) - l_m \cos\theta_S \\ z_{m,n} &= l_m \sin\theta_S \end{aligned} \quad (1)$$

where $\psi_c$ ($\psi_c \triangleq \frac{2\pi}{N-1}$) is the angular distance between the centers of two planks (Fig. 1) and $l_m$ ($m = 1, ..., M$) is the distance between the first element and the $m$-th one, being $l_1 = 0$ and $l_M = l$.

The generation of $B$ pencil beams (Fig. 2) pointing along different elevation directions, $\{(\theta^{(b)}, \phi^{(b)}); b = 1, ..., B\}$, is obtained by controlling the amplitude and the phase weights of the *CPA* excitations by means of the fully *DBF* architecture sketched in Fig. 3. More in detail, the signal received by each antenna is digitized with an *ADC* and it is weighted by the set of amplitude weights, $\boldsymbol{\alpha}$ ($\boldsymbol{\alpha} \triangleq \{\alpha_{m,n}^{(b)}; m = 1, ..., M; n = 1, ..., N; b = 1, ..., B\}$), and phase delays, $\boldsymbol{\varphi}$ ($\boldsymbol{\varphi} \triangleq \{\varphi_{m,n}^{(b)}; m = 1, ..., M; n = 1, ..., N; b = 1, ..., B\}$), that are combined to afford the $B$ independent patterns, each $b$-th ($b = 1, ..., B$) beam pointing towards the user-defined angular direction $(\theta^{(b)}, \phi^{(b)})$. The electromagnetic far-field pattern of the $b$-th beam ($b = 1, ..., B$) radiated by the *CPA* is equal to

$$\mathbf{E}^{(b)}(\theta, \phi) = \mathbf{e}(\theta, \phi) \times F^{(b)}(\theta, \phi) \quad (2)$$

where $\mathbf{e}(\theta, \phi)$ is the embedded/active-element pattern [23][24] and $F^{(b)}(\theta, \phi)$ is the array factor given by

$$\begin{aligned} F^{(b)}(\theta, \phi) = \sum_{n=1}^{N} \sum_{m=1}^{M} \alpha_{m,n}^{(b)} \exp \Big\{ j\tfrac{2\pi}{\lambda} \big[ &x_{m,n} \left( \sin\theta \cos\phi - \sin\theta^{(b)} \cos\phi^{(b)} \right) + \\ &y_{m,n} \left( \sin\theta \sin\phi - \sin\theta^{(b)} \sin\phi^{(b)} \right) + z_{m,n} \left( \cos\theta - \cos\theta^{(b)} \right) \big] + j\varphi_{m,n}^{(b)} \Big\}, \end{aligned} \quad (3)$$

$\lambda$ being the free-space wavelength at the working frequency $f_0$.

Because of the modularity of the *CPA* at hand and that pointing the mainlobe towards the $b$-th ($b = 1, ..., B$) direction, $\left(\theta^{(b)}, \phi^{(b)}\right)$, means analytically adding to the $(m, n)$-th ($m = 1, ..., M$; $n = 1, ..., N$) digital phase coefficient, $\varphi_{m,n}^{(b)}$, a shift value $\Delta\varphi_{m,n}^{(b)}$

$$\Delta\varphi_{m,n}^{(b)} = -\frac{2\pi}{\lambda}\left(x_{m,n}\sin\theta^{(b)}\cos\phi^{(b)} + y_{m,n}\sin\theta^{(b)}\sin\phi^{(b)} + z_{m,n}\cos\theta^{(b)}\right), \quad (4)$$

the degrees of freedom (*DoF*s) in designing such a *CPA* architecture turn out to be only those of a single plank, namely the positions of the $M$-elements sparse linear array of the plank and the corresponding amplitude, $\boldsymbol{\alpha}^{(b)} = \left\{\alpha_m^{(b)}; m = 1, ..., M\right\}$ ($b = 1, ..., B$), and phase, $\boldsymbol{\varphi}^{(b)} = \left\{\varphi_m^{(b)}; m = 1, ..., M\right\}$ ($b = 1, ..., B$), excitation coefficients, being $\alpha_{m,n}^{(b)} = \alpha_m^{(b)}$ and $\varphi_{m,n}^{(b)} = \varphi_m^{(b)} - \frac{2\pi}{\lambda}\left(x_{m,n}\sin\theta_S + z_{m,n}\cos\theta_S\right)$ ($m = 1, ..., M$; $n = 1, ..., N$). The design of the *CPA* is then cast as the synthesis of the linear sparse array, which composes the vertical plank, that simultaneously generates $B$ beams. Mathematically, it can be stated as follows:

> *Multi-Beam Sparse Array Synthesis Problem* (*MBSASP*) - Given a set of $B$ reference beam patterns, $\{\widehat{F}^{(b)}(\theta); b = 1, ..., B\}$, radiated by a reference *FP* linear array of $I$ elements uniformly spaced by $d$ and pointing towards $B$ directions along the elevation plane (Fig. 2), $\{\theta^{(b)}; b = 1, ..., B\}$, determine the corresponding $M$-element ($M < I$) maximally sparse arrangement and the set of $B$ amplitude $\boldsymbol{\alpha}^{(b)}$ and phase $\boldsymbol{\varphi}^{(b)}$ excitation coefficients ($b = 1, ..., B$), such that $M$ is minimum and the multi-beam pattern matching constraint
>
> $$\frac{\sum_{b=1}^{B}\sum_{k=1}^{K}\left|\widehat{F}^{(b)}(\theta_k) - F^{(b)}\left(\theta_k; \boldsymbol{\alpha}^{(b)}, \boldsymbol{\varphi}^{(b)}\right)\right|^2}{B \times K} < \epsilon \quad (5)$$
>
> is satisfied, $\{\theta_k; k = 1, ..., K\}$ being the set of $K$ sampling directions along the elevation plane, while $\epsilon$ is the user-defined parameter controlling the degree of accuracy of the patterns matching.

To solve the *MBSASP*, let us consider a coordinate system with the $\xi$-axis along the plank column (Figs. 1-2) so that the corresponding broadside direction, $\theta' = 90$ [deg], coincides with the slant angle $\theta = \theta_S$ of the reference coordinate system (Fig. 2) and the non-uniform positions of the $M$ elements of the plank, $\{\xi_m; m = 1, ..., M\}$, belong to a set of $Q$ ($Q \gg M$) user-defined candidate locations, $\{\xi_q; q = 1, ..., Q\}$, of a uniform lattice [see for instance Fig. 6(a)]. Accordingly, the synthesis of the $B$ sets of complex excitation weights $\boldsymbol{\gamma}^{(b)} \triangleq \{\gamma_q^{(b)} = \delta_{qm}\alpha_m^{(b)}e^{j\varphi_m^{(b)}}; q = 1, ..., Q; m = 1, ..., M\}$, $\delta_{qm}$ being the Kronecker function ($\delta_{qm} = 1$ if $\xi_q = \xi_m$ and $\delta_{qm} = 0$, otherwise), is mathematically formulated as

$$\boldsymbol{\gamma}_{\ell_0}^{(b)} = \arg\left\{\min_{\boldsymbol{\gamma}^{(b)}} \|\boldsymbol{\gamma}^{(b)}\|_{\ell_0}\right\} \quad (6)$$

subject to

$$\widehat{\mathbf{F}}^{(b)} - \mathbf{A}\boldsymbol{\gamma}^{(b)} = \boldsymbol{\eta}^{(b)} \quad b = 1, ..., B \quad (7)$$

where $\mathbf{A}$ is the steering matrix ($\mathbf{A} \triangleq \{a_{kq} = e^{j\frac{2\pi}{\lambda}\xi_q \cos\theta_k'}; k = 1, ..., K; q = 1, ..., Q\}$), $\widehat{\mathbf{F}}^{(b)}$ ($\widehat{\mathbf{F}}^{(b)} \triangleq \{\widehat{F}^{(b)}(\theta_k'); k = 1, ..., K\}$) is the set of $K$ samples of the $b$-th ($b = 1, ..., B$) reference pattern, while $\boldsymbol{\eta}^{(b)} = \{\eta_k^{(b)}; k = 1, ..., K\}$ is the $b$-th ($b = 1, ..., B$) noise vector whose entries are zero-mean Gaussian complex error values with variance $\sigma$ proportional to $\epsilon$.

The *MT-BCS* strategy [29] is then adopted and customized to deal with the ill-posed/ill-conditioned problem in (6) by statistically correlating the $B$ sets of array excitations. This implies to enforce the linear array having non-null excitations at the same positions of the $Q$-locations uniform lattice so that the sparsest array layout (i.e., the minimum value of $M$) is retrieved. More specifically, each $b$-th ($b = 1, ..., B$) constraint in (7) is coded into 2 real-valued tasks[1]

$$\begin{cases} \widetilde{\mathbf{F}}_{\mathcal{R}}^{(b)} - \widetilde{\mathbf{A}}\boldsymbol{\gamma}_{\mathcal{R}}^{(b)} = \widetilde{\boldsymbol{\eta}}_{\mathcal{R}}^{(b)} \\ \widetilde{\mathbf{F}}_{\mathcal{I}}^{(b)} - \widetilde{\mathbf{A}}\boldsymbol{\gamma}_{\mathcal{I}}^{(b)} = \widetilde{\boldsymbol{\eta}}_{\mathcal{I}}^{(b)} \end{cases} \quad (8)$$

where $\boldsymbol{\gamma}_{\mathcal{R}}^{(b)} \triangleq \mathcal{R}\{\boldsymbol{\gamma}^{(b)}\}$ and $\boldsymbol{\gamma}_{\mathcal{I}}^{(b)} \triangleq \mathcal{I}\{\boldsymbol{\gamma}^{(b)}\}$, $\mathcal{R}\{\cdot\}$ and $\mathcal{I}\{\cdot\}$ being the real part and the imaginary one, respectively, $\widetilde{\mathbf{F}}_{\mathcal{R}}^{(b)} \triangleq \left[\mathcal{R}\{\widehat{\mathbf{F}}_{\mathcal{R}}^{(b)}\}, \mathcal{I}\{\widehat{\mathbf{F}}_{\mathcal{R}}^{(b)}\}\right]$ and $\widetilde{\mathbf{F}}_{\mathcal{I}}^{(b)} \triangleq \left[\mathcal{R}\{\widehat{\mathbf{F}}_{\mathcal{I}}^{(b)}\}, \mathcal{I}\{\widehat{\mathbf{F}}_{\mathcal{I}}^{(b)}\}\right]$ are two real-valued vectors such that $\widetilde{\mathbf{F}}_{\mathcal{R}}^{(b)} + \widetilde{\mathbf{F}}_{\mathcal{I}}^{(b)} = \left[\mathcal{R}\{\widehat{\mathbf{F}}^{(b)}\}, \mathcal{I}\{\widehat{\mathbf{F}}^{(b)}\}\right]$. Moreover, $\widetilde{\boldsymbol{\eta}}_{\mathcal{R}}^{(b)}$ and $\widetilde{\boldsymbol{\eta}}_{\mathcal{I}}^{(b)}$ are also real-valued vectors such that $\widetilde{\boldsymbol{\eta}}_{\mathcal{R}}^{(b)} + \widetilde{\boldsymbol{\eta}}_{\mathcal{I}}^{(b)} = \left[\mathcal{R}\{\boldsymbol{\eta}^{(b)}\}, \mathcal{I}\{\boldsymbol{\eta}^{(b)}\}\right]$, and $\widetilde{\mathbf{A}} \triangleq$

---

[1] The problem constraints are transformed in a real-valued form to enable the use of *CS*-based state-of-the-art algorithms [34].

$[\mathcal{R}\{\mathbf{A}\}, \mathcal{I}\{\mathbf{A}\}]$ is a $2K \times Q$ real-valued matrix.

The sparsest weights vectors $\boldsymbol{\gamma}_{\mathcal{R}}^{(b)}$ and $\boldsymbol{\gamma}_{\mathcal{I}}^{(b)}$ ($b = 1, ..., B$) are then derived by maximizing, component-by-component, the *a-posteriori* probability of having the $\boldsymbol{\gamma}^{(b)}$ coefficients in correspondence with the set of reference pattern samples $\widehat{\mathbf{F}}^{(b)}$

$$\begin{cases} \boldsymbol{\gamma}_{\mathcal{R}}^{(b)} = \arg\left[\max_{\widetilde{\boldsymbol{\gamma}}_{\mathcal{R}}^{(b)}} \mathcal{P}\left(\widetilde{\boldsymbol{\gamma}}_{\mathcal{R}}^{(b)} \mid \widetilde{\mathbf{F}}_{\mathcal{R}}^{(b)}\right)\right] \\ \boldsymbol{\gamma}_{\mathcal{I}}^{(b)} = \arg\left[\max_{\widetilde{\boldsymbol{\gamma}}_{\mathcal{I}}^{(b)}} \mathcal{P}\left(\widetilde{\boldsymbol{\gamma}}_{\mathcal{I}}^{(b)} \mid \widetilde{\mathbf{F}}_{\mathcal{I}}^{(b)}\right)\right] \end{cases}. \quad (9)$$

It turns out that [29]

$$\begin{cases} \boldsymbol{\gamma}_{\mathcal{R}}^{(b)} = \left[diag\left(\widetilde{\mathbf{a}}\right) + \widetilde{\mathbf{A}}^T\widetilde{\mathbf{A}}\right]^{-1}\widetilde{\mathbf{A}}^T\widetilde{\mathbf{F}}_{\mathcal{R}}^{(b)} \\ \boldsymbol{\gamma}_{\mathcal{I}}^{(b)} = \left[diag\left(\widetilde{\mathbf{a}}\right) + \widetilde{\mathbf{A}}^T\widetilde{\mathbf{A}}\right]^{-1}\widetilde{\mathbf{A}}^T\widetilde{\mathbf{F}}_{\mathcal{I}}^{(b)} \end{cases} \quad (10)$$

where $\widetilde{\mathbf{a}}$ is the shared hyper-parameter vector determined with the *RVM* solver [38].

Finally, the $b$-th ($b = 1, ..., B$) set of complex excitations affording the desired $b$-th beam is yielded as follows

$$\boldsymbol{\gamma}^{(b)} = \boldsymbol{\gamma}_{\mathcal{R}}^{(b)} + j\boldsymbol{\gamma}_{\mathcal{I}}^{(b)}. \quad (11)$$

# 3 Numerical Results

This section has a twofold objective. On the one hand, to assess the effectiveness of a sparse *CPA* architecture for *ATC* applications, on the other, the validation of the proposed *BCS*-based multi-beam synthesis method.

In the benchmark scenario, the *CPA* has been required to generate $B = 7$ beams pointing at different elevation angles such that each beam intersects at $-3$ [dB] the left and the right neighboring beams [Fig. 4($b$)] to ensure a $-3$ [dB] power coverage within an angular range of $\Theta_s = 40$ [deg]. To span the elevation range $\theta \in [50 : 90]$ [deg], the slant angle has been set to $\theta_S = 70$ [deg] [36][37], the horizon being at $\theta = 90$ [deg]. Therefore, the $B$ steering angles in the plank coordinate system, $\{\theta'^{(b)}; b = 1, ..., B\}$, have been set as in Tab. I, where the corresponding $-3dB_{left}$ and $-3dB_{right}$ power pattern intersection points at $-3$ [dB] and the half power beamwidth (*HPBW*) are reported, as well.

As a starting point for the *MT-BCS* synthesis of the single plank, the linear configuration of $I = 22$ uniformly-spaced ($d = \frac{\lambda}{2} \Rightarrow l = 10.5\,\lambda$) elements in Fig. 4(*a*) has been used as reference, while the normalized power patterns [Fig. 4(*b*)] radiated by the corresponding (reference) *FP-CPA*, composed by $N = 204$ columns spaced by $d_c = \frac{\lambda}{2}$ for a total of $P = N \times I = 4488$ elements, have been synthesized by tapering the array excitations with a Taylor distribution [39] having $SLL = -30$ [dB] and $\bar{n} = 6$. These choices resulted in a cone geometry with minor (major) radius equal to $r \simeq 15.92\,\lambda$ ($R \simeq 19.51\,\lambda$). Successively, multiple runs of the *MT-BCS* code have been performed by varying the lattice (i.e., the number of $Q$ partitions of $l$: $20 \leq \frac{Q}{I} \leq 40$) and the reference pattern sampling (i.e., $K$: $1 \leq \frac{K}{I} \leq 3$) as well as the *MT-BCS* control parameters according to the guidelines in [29]-[31] (i.e., $\sigma \in [10^{-5}, 10^{-2}]$, $\beta_1 \in [10^{-1}, 10^4]$, and $\beta_2 \in [5 \times 10^{-1}, 5 \times 10^2]$, $\beta_1$ and $\beta_2$ being the *MT* hyper-priors [30]) to explore the achievable trade-offs between pattern matching performance and array sparseness. The degree of optimality of the synthesized layouts has been quantified with the multi-beam power pattern matching error, $\chi$, defined as $\chi \triangleq \frac{1}{B} \sum_{b=1}^{B} \chi^{(b)}$ being

$$\chi^{(b)} \triangleq \frac{\int_0^\pi \left| \left|\widehat{F}^{(b)}(\theta)\right|^2 - \left|F^{(b)}(\theta)\right|^2 \right| d\theta}{\int_0^\pi \left|\widehat{F}^{(b)}(\theta)\right|^2 d\theta} \qquad (12)$$

the single $b$-th ($b = 1, ..., B$) beam error. The arising Pareto front of the solutions in the $M - \chi$ plane is shown in Fig. 5. As it can be observed, the error drops down the value of $\chi = 6.1 \times 10^{-3}$ with at least $M = 16$ elements ($Q = 700$, $K = 44$, $\sigma = 10^{-5}$, $\beta_1 = 10^{-1}$, and $\beta_2 = 5 \times 10^{-1}$). In this latter case, the array layout is characterized by an aperture of length $l\rfloor_{M=16} = 10.33\,\lambda$ and a maximum (minimum) inter-element spacing equal to $d_m^{\max} = 0.706\lambda$ ($d_m^{\min} = 0.646$) [Fig. 6(*a*)]. The $B = 7$ radiated beam patterns are compared to the reference ones in Fig. 6(*b*) to give an insight on the pattern matching accuracy. For completeness, the $B$ sets of the optimized complex excitations, $\{\boldsymbol{\gamma}^{(b)}; b = 1, ..., B\}$, are shown in Fig. 6(*c*), while the values of the matching error for each $b$-th ($b = 1, ..., B$) beam, $\chi^{(b)}$, are given in Tab. II together with the main pattern descriptors [i.e., the side lobe level (*SLL*), the peak directivity (*D*), and the half power beamwidth ($HPBW$)]. It is worth noticing that the *MT-BCS* design faithfully matches the whole set of reference patterns (e.g., the maximum degradations of the *SLL*, of *D*, and of $HPBW$ being 1.21 [dB], 0.04 [dBi], and 0.05 [deg], respectively) despite a reduction of 27.3 % of the elements of the *FP* array. For illustrative purposes, the inset of Fig. 6(*b*) shows the case with the worst matching error in Tab. II (i.e., $b = B = 7$).

Once the sparse plank has been synthesized, the *CPA* has been assembled by subdividing the lateral surface of the cone into $S$ vertical sectors, each composed by $N_c$ contiguous planks. The $N_c$ planks, belonging to the $s$-th ($s = 1, ..., S$) vertical sector, are responsible of generating $B$ beams pointing towards the desired $B$ directions along the elevation plane, $\{\theta^{(b)}; b = 1, ..., B\}$, while having the same azimuth angle [i.e., $\phi^{(b)} = \phi_s$ ($b = 1, ..., B$) - Figs. 7(*a*)-7(*c*)]. To assess the performance of the $3D$ conical array in focusing the beam along elevation and azimuth, different *CPA* configurations have been taken into account by setting the angular width of the vertical sector to $\psi = 30$ [deg] [Fig. 7(*a*)], $\psi = 60$ [deg] [Fig. 7(*b*)], and $\psi = 90$ [deg] [Fig. 7(*c*)], which means an architecture of $S = 12$ sectors with $N_c = 17$ planks [Fig. 7(*a*) and Fig. 7(*g*)], $S = 6$ sectors with $N_c = 34$ planks [Fig. 7(*e*) and Fig. 7(*h*)], and $S = 4$ sectors with $N_c = 51$ planks [Fig. 7(*e*) and Fig. 7(*i*)], respectively. Figure 8 shows in a color-map representation the normalized power patterns in the $(v, w)$-plane ($v \triangleq \sin\theta \sin\phi$, $w \triangleq \cos\theta$, being $\theta \in [0 : 180]$ [deg] and $\phi \in [-90 : 90]$ [deg]) of a subset (i.e., $b = \{1, 3, 5, 7\}$) of the $B = 7$ beams radiated by the $\psi = 30$ [deg] sector *CPA*. The plots refer to three different frequencies within the *L-Band*, which is a typical frequency range reserved for aeronautical radio-navigation/radio-localization and, in particular, for Primary Surveillance Radar (*PRS*) applications [1]. More specifically, the frequencies $f_{\min} = 1.215$ [GHz] [Figs. 8(*a*)-8(*d*)], $f_0 = 1.282$ [GHz] [Figs. 8(*e*)-8(*h*)], and $f_{\max} = 1.350$ [GHz] [Figs. 8(*i*)-8(*n*)] have been analyzed and the performance of the sparse *CPA*s [Figs. 7(*g*)-7(*i*)] with respect to the reference *FP* ones [Figs. 7(*d*)-7(*f*)] have been evaluated still with the pattern matching metric in (12), but now considering the two angular variables $(v, w)$ (i.e., $\chi^{(b)} \triangleq \frac{\int_{v^2+w^2 \leq 1} \left||\widehat{F}^{(b)}(v,w)|^2 - |F^{(b)}(v,w)|^2\right| dvdw}{\int_{v^2+w^2 \leq 1} |\widehat{F}^{(b)}(v,w)|^2 dvdw}$). The behavior of $\chi^{(b)}$ versus the beam number ($b = 1, ..., B$) at the selected frequencies can be inferred by the plots in left column of Fig. 9, which refer to the $\psi = 30$ [deg] [Fig. 9(*a*)], the $\psi = 60$ [deg] [Fig. 9(*c*)], the $\psi = 90$ [deg] [Fig. 9(*e*)] sectorized *CPA*, respectively. Generally, the error values are in the order of $\chi^{(b)} \approx 10^{-4}$ and slightly increase ($\chi^{(b)} \approx 10^{-3}$) only for the border beams (i.e., $b = 1$ and $b = B$) at the higher frequency (i.e., $f_{\max} = 1.350$ [GHz]). To give the interested readers some insights on the distribution of the error done in approximating the reference pattern within the (v,w)-plane, a local mismatch index, $\Delta$, has been defined as $\Delta \triangleq \left| \left| \widehat{F}^{(b)}(v,w) \right|^2 - \left| F^{(b)}(v,w) \right|^2 \right| / \left| \widehat{F}^{(b)}(v,w) \right|^2$ and it has been computed for the worst cases having the greater values of $\chi^{(b)}$ (i.e., $b = 1$ @ $f_{\max} = 1.350$ [GHz]) [Fig. 9 - right column]. As it can be inferred, the most significant deviations from the reference pattern turn out to be close to the $(\theta, \phi) = (180, 0)$ [deg] [$\rightarrow (v, w) = (0, -1)$] angular direction, that is, far away from the main-beam in the low sidelobe region [Fig. 8(*i*)].

Next, the main pattern descriptors [i.e., the *HPBW* along the azimuth ($HPBW_{AZ}$) and elevation ($HPBW_{EL}$), the *SLL*, the *SLL* in the elevation plane ( $SLL_{EL} \triangleq SLL\rfloor_{v=0}$), and the peak directivity (*D*)] of the different sparse *CPA* configurations have been analyzed. As a representative example of the whole set of results, the discussion will be focused on the central beam (i.e., $b = 4$). Figure 10(*a*) shows the behaviors of $HPBW_{AZ}$ and $HPBW_{EL}$ versus the sector width $\psi$ of the *CPA* architecture. As expected, there is an unavoidably beam broadening effect when increasing the operation frequency and the $HPBW_{AZ}$ reduces of almost one third widening the angular width of the vertical sector from $\psi = 30$ [deg] up to $\psi = 90$ [deg]. Concerning the values of *SLL* and *D*, which are reported in Fig. 10(*b*), it turns out that *D* increases with the sector width due to the larger size of the aperture that radiates the beam, but the same holds true for the *SLL* due to the high sidelobes in the azimuth plane since the *CPA* sector behaves as a uniform array along such a plane (i.e., all planks as well as the element excitations are equal and without tapering along the azimuth plane). However, the azimuth-plane sidelobes can be easily lowered by exploiting, for instance, the pattern multiplication strategy [23][24]. Accordingly, a Taylor taper [39] with $SLL = -30$ [dB] and $\bar{n} = 4$ has been applied to the amplitudes of the $N_c = 17$ planks of the sparse *CPA* assembled with $\psi = 30$ [deg] vertical sectors. As expected, the plots of the power patterns radiated at the central frequency $f_0$ for the $b = \{1, 3, 5, 7\}$ beams in Fig. 11 do not present the high sidelobes along the azimuth plane of the corresponding ones in Figs. 8(*e*)-8(*h*). Indeed, the *SLL* now turns out to be $SLL \leq -26.78$ [dB], which is a value very close to the reference Taylor one.

Finally, the reliability of the proposed sparse *CPA* modular architecture and its robustness against the non-idealities of real arrays have been assessed. Towards this end, a square-ring microstrip antenna [Fig. 12(*a*)], suitable for wide angle scanning [40] and resonating in the *L*-band, has been chosen as elementary radiator of the array. To include the mutual coupling effects of the real array structure, the corresponding element pattern, $\mathbf{e}(\theta, \phi)$ ($\mathbf{e}(\theta, \phi) \neq \mathbf{1}$, $\mathbf{1}$ being the element pattern of the ideal-isotropic radiator), has been set to the embedded element pattern of the central element of a neighborhood of $5 \times 5$ identical square-ring microstrip antennas conformal to the *CPA* support [Fig. 12(*b*)]. For illustrative purposes, the *3D* plots of the embedded pattern at the frequencies of interest are reported: $f_{\min} = 1.215$ [GHz] [Fig. 12(*c*)], $f_0 = 1.282$ [GHz] [Fig. 12(*d*)], and $f_{\max} = 1.350$ [GHz] [Fig. 12(*e*)]. To analyze the radiation performance, Figure 13 compares, along the elevation plane, the power patterns radiated at $f_0 = 1.282$ [GHz] by the real and the ideal (i.e., $\mathbf{e}(\theta, \phi) = \mathbf{1}$) sparse *CPA*s in the $\psi = 30$ [deg] [Fig. 13(*a*)], $\psi = 60$ [deg] [Fig. 13(*b*)], and $\psi = 90$ [deg] [Fig. 13(*c*)] configurations. As it can be observed, the real and the ideal curves are substantially identical in the main beam region, while some negligible differences appear in the far sidelobe region.

## 4  Conclusions

The design of a sparse *CPA*, which generates multiple beams pointing along different elevation directions, to be used as receiver for next-generation *ATC* radar systems has been carried out. Thanks to a modular structure composed by vertical modules/planks consisting of a sparsely linearly-arranged set of radiating elements, the array, equipped with a fully *DBF* network to simultaneously generate multiple beams on receive, has been conceived to minimize the architecture complexity and the number of radiating elements as well as *ADC*s. The *CPA* synthesis has been carried out by means of a customized implementation of the *MT-BCS*-based method and it is aimed at jointly optimizing the positions of the plank radiators and the set of complex excitations for generating the multiple beams.

From the numerical assessment with ideal as well as real antenna models, the following main outcomes can be drawn:

- a sparse *CPA* with $27.3$ % less elements than the reference *FPA*, while guaranteeing the same radiation performance, has been synthesized thanks to the proposed *CS*-based method;

- the effectiveness of the *MT-BCS* synthesis has been proved also in solving array design problems with multiple concurrent tasks such as the one here addressed and concerned with the simultaneous generation of multiple beams with the same sparse physical architecture;

- the modular structure of the *CPA* along the elevation plane allows the designer to choose the best trade-off in terms of resolution, radar range, and tracking directions subject to the requirements on the *ATC* radar at hand.

Future research activities, beyond the scope of this paper, will deal with innovative unconventional *CPA* architectures that exploit sparsity on both elevation and azimuth as well as other sizes/shapes of the planks to further address the easy-manufacturing and cost-reduction issues of future-generation multi-function radar systems. Of course, the extension of current and advances *CPA* geometries to other frequency bands will be object of more application-oriented research tracks.

# Acknowledgements

This work benefited from the networking activities carried out within the Project Project "CYBER-PHYSICAL ELECTROMAGNETIC VISION: Context-Aware Electromagnetic Sensing and Smart Reaction (EMvisioning)" (Grant no. 2017HZJXSZ)" funded by the Italian Ministry of Education, University, and Research under the PRIN2017 Program (CUP: E64I19002530001). Moreover, it benefited from the networking activities carried out within the Project "SPEED" (Grant No. 61721001) funded by National Science Foundation of China under the Chang-Jiang Visiting Professorship Program, the Project 'Inversion Design Method of Structural Factors of Conformal Load-bearing Antenna Structure based on Desired EM Performance Interval' (Grant no. 2017HZJXSZ) funded by the National Natural Science Foundation of China, and the Project 'Research on Uncertainty Factors and Propagation Mechanism of Conformal Loab-bearing Antenna Structure' (Grant No. 2021JZD-003) funded by the Department of Science and Technol-ogy of Shaanxi Province within the Program Natural Science Basic Research Plan in Shaanxi Province. A. Massa wishes to thank E. Vico for her never-ending inspiration, support, guidance, and help.

[18] W. Dong, Z. Xu, X. Liu, L. Wang, and S. Xiao, "Modular subarrayed phased-array design by means of iterative convex relaxation optimization," *IEEE Antennas Wireless Propag. Lett.*, vol. 18, no. 3, pp. 447-451, Mar. 2019.

[19] Y. Ma, S. Yang, Y. Chen, S. Qu, and J. Hu, "Pattern synthesis of 4-D irregular antenna arrays based on maximum-entropy model," *IEEE Trans. Antennas Propag.*, vol. 67, no. 5, pp. 3048-3057, May 2019.

[20] P. Rocca, N. Anselmi, A. Polo, and A. Massa, "An irregular two-sizes square tiling method for the design of isophoric phased arrays," *IEEE Trans. Antennas Propag.*, vol. 68, no. 6, pp. 4437-4449, Jun. 2020.

[21] P. Rocca, N. Anselmi, A. Polo, and A. Massa, "Modular design of hexagonal phased arrays through diamond tiles," *IEEE Trans. Antennas Propag.*, vol. 68, no. 5, pp. 3598-3612, May 2020.

[22] B. Rupakula, A. H. Aljuhani, and G. M. Rebeiz, "Limited scan-angle phased-arrays using randomly-grouped subarrays and reduced number of phase-shifters," *IEEE Trans. Antennas Propag.*, vol. 68, no. 1, pp. 70-80, Jan. 2020.

[23] R. L. Haupt, *Antenna Arrays - A Computation Approach.* Hoboken, NJ, USA: Wiley, 2010.

[24] R. J. Mailloux, *Phased Array Antenna Handbook* (3rd Ed.). Boston, MA, USA: Artech House, 2018.

[25] A. Trucco and V. Murino, "Stochastic optimization of linear sparse arrays," *IEEE J. Oceanic Eng.*, vol. 24, no.3, pp. 291-299, Jul. 1999.

[26] Y. Liu, Q. H. Liu, and Z. Nie, "Reducing the number of elements in the synthesis of shaped-beam pattern by the forward-backward matrix pencil method," *IEEE Trans. Antennas Propag.*, vol. 58, no. 2, pp. 604-608, Feb. 2010.

[27] S. K. Goudos, K. Siakavara, T. Samaras, E. E. Vafiadis, and J. N. Sahalos, "Sparse linear array synthesis with multiple constraints using differential evolution with strategy adaptation," *IEEE Antennas Wireless Propag. Lett.*, vol. 10, pp. 670-673, 2011.

# FIGURE CAPTIONS

- **Figure 1.** Sketch of the sparse *CPA* geometry.

- **Figure 2.** Sketch of the transversal section of the *CPA* and of the radiation mechanism.

- **Figure 3.** Logical scheme of the *DBF* network of the single plank that generates $B$ independent radiation beams.

- **Figure 4.** *Numerical Assessment* ($I = 22$, $d = \frac{\lambda}{2}$, $B = 7$) - Plot of (*a*) the positions of the elements of the *FP* reference linear array and of (*b*) the $B$ radiated (reference) power patterns.

- **Figure 5.** *Numerical Assessment* ($I = 22$, $d = \frac{\lambda}{2}$, $B = 7$) - Representative points in the ($M$,$\chi$)-plane of the *CS*-synthesized plank layouts and the corresponding Pareto front.

- **Figure 6.** *Numerical Assessment* ($M = 16$, $d = \frac{\lambda}{2}$, $B = 7$) - Plot of (*a*) the *MT-BCS* synthesized layout of the plank and of (*b*) the $B$ radiated power patterns when the sparse *CPA* is fed with (*c*) the corresponding sets of complex excitations.

- **Figure 7.** *Numerical Assessment* ($M = 16$, $d = \frac{\lambda}{2}$, $B = 7$) - *CPA* configurations when partitioning the lateral surface of the cone in vertical sectors with (*a*)(*d*)(*g*) $\psi = 30$ [deg] ($\to S = 12$ and $N_c = 17$), (*b*)(*e*)(*h*) $\psi = 60$ [deg] ($\to S = 6$ and $N_c = 34$), and (*c*)(*f*)(*i*) $\psi = 90$ [deg] ($\to S = 4$ and $N_c = 51$) widths by using (*d*)-(*f*) the *FP* and (*g*)-(*i*) the sparse planks.

- **Figure 8.** *Numerical Assessment* (Sparse *CPA*, $M = 16$, $d = \frac{\lambda}{2}$, $B = 7$, $\psi = 30$ [deg]) - Plot of the power pattern in the (*v*,*w*)-plane of the (*a*)(*e*)(*i*) $b = 1$, (*b*)(*f*)(*j*) $b = 3$, (*c*)(*g*)(*k*) $b = 5$, and (*d*)(*h*)(*l*) $b = 7$ beams at (*a*)-(*d*) $f_{min} = 1.215$ [GHz], (*e*)-(*h*) $f_0 = 1.282$ [GHz], and (*i*)-(*l*) $f_{max} = 1.350$ [GHz].

- **Figure 9.** *Numerical Assessment* (Sparse *CPA*, $M = 16$, $d = \frac{\lambda}{2}$, $B = 7$) - Plot of (*a*)(*c*)(*e*) the pattern matching error, $\chi^{(b)}$, versus the beam index, $b$ ($b = 1, ..., B$), and of (*b*)(*d*)(*f*) the local mismatch index, $\Delta$, in the (*v*,*w*)-plane for the $b = 1$ beam at $f_{max} = 1.350$ [GHz] in correspondence with the *CPA* configurations with vertical sector width equal to (*b*) $\psi = 30$ [deg], (*d*) $\psi = 60$ [deg], and (*f*) $\psi = 90$ [deg].

- **Figure 10.** *Numerical Assessment* (Sparse *CPA*, $M = 16$, $d = \frac{\lambda}{2}$, $b = 4$, $B = 7$) - Behaviour of (*a*) $HPBW_{AZ}$ and $HPBW_{EL}$ and of (*b*) $SLL$ and $D$ versus the vertical sector width, $\psi$.

- **Figure 11.** *Numerical Assessment* (Sparse *CPA*, $M = 16$, $d = \frac{\lambda}{2}$, $B = 7$, $f_0 = 1.282$ [GHz], $\psi = 30$ [deg]) - Plot of the power pattern in the (*v*,*w*)-plane of the (*a*) $b = 1$, (*b*) $b = 3$, (*c*) $b = 5$, and (*d*) $b = 7$ beams when applying a Taylor tapering with $SLL = -30$ [dB] and $\bar{n} = 4$ to the excitations of the $N_c = 17$ planks in the azimuth plane.

- **Figure 12.** *Numerical Assessment* - Sketch of (a) the model of the elementary radiator and of (*a*) the $5 \times 5$ conformal neighborhood used to compute (*c*)(*d*)(*e*) the embedded element pattern at (*c*) $f_{min} = 1.215$ [GHz], (*d*) $f_0 = 1.282$ [GHz], and (*e*) $f_{max} = 1.350$ [GHz].

- **Figure 13.** *Numerical Assessment* (Sparse *CPA*, $M = 16$, $d = \frac{\lambda}{2}$, $B = 7$, $f_0 = 1.282$ [GHz]) - Plot along the $\phi = 0.0$ [deg] cut of the normalized power patterns radiated by the CPA configurations with vertical sector width equal to (*a*) $\psi = 30$ [deg], (*b*) $\psi = 60$ [deg], and (*c*) $\psi = 90$ [deg].

# TABLE CAPTIONS

- **Table I.** *Numerical Assessment* (*FP-CPA*, $I = 22$, $d = \frac{\lambda}{2}$, $B = 7$) - Steering angles and *HPBW* values.

- **Table II.** *Numerical Assessment* ($I = 22$, $M = 16$, $d = \frac{\lambda}{2}$, $B = 7$) - Pattern indexes and pattern matching errors.

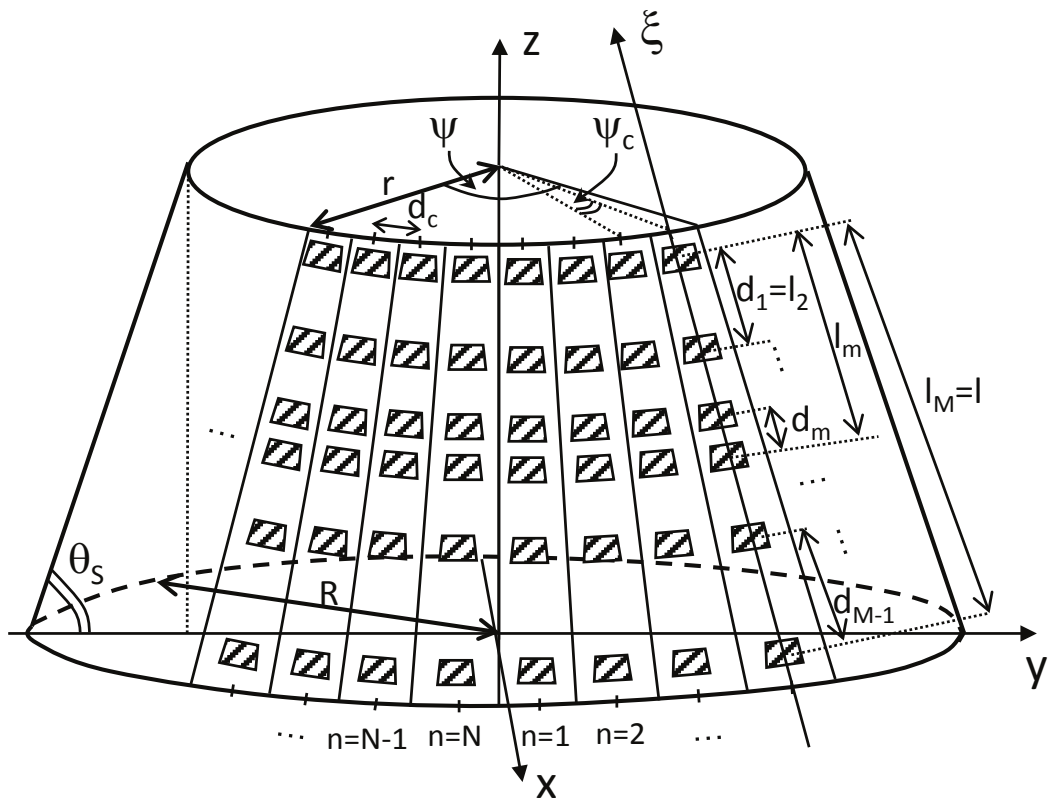

**Fig. 1 - P. Rocca *et al.*,** "Conical Multi-Beam Phased Arrays for ..."

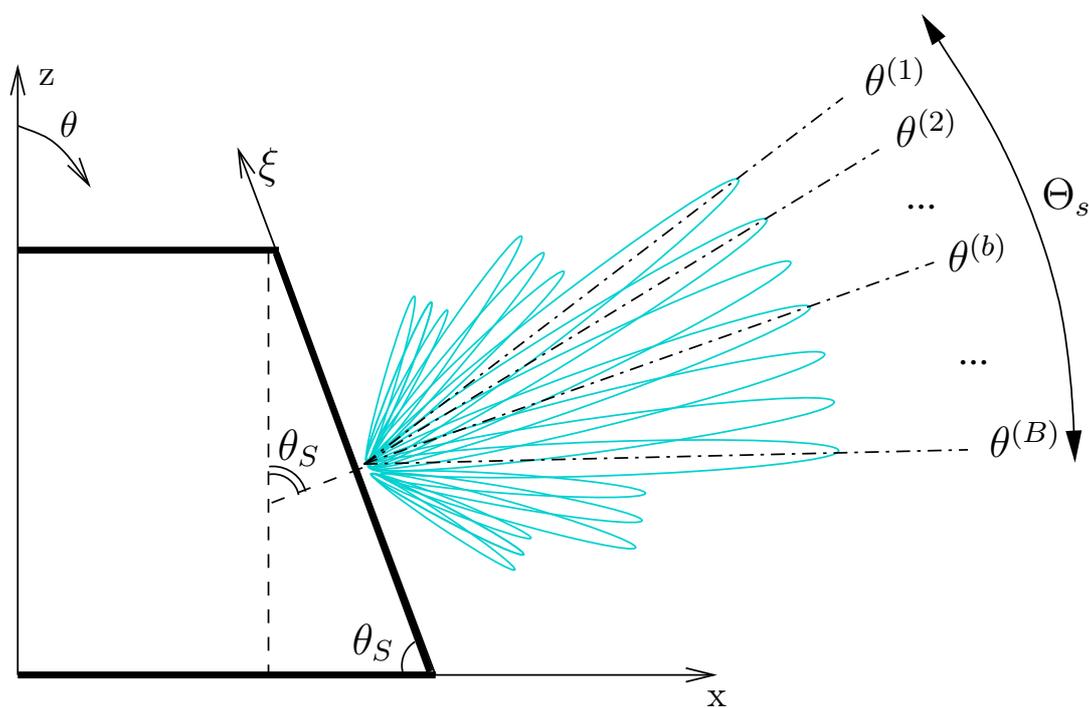

**Fig. 2** - P. Rocca *et al.*, "Conical Multi-Beam Phased Arrays for ..."

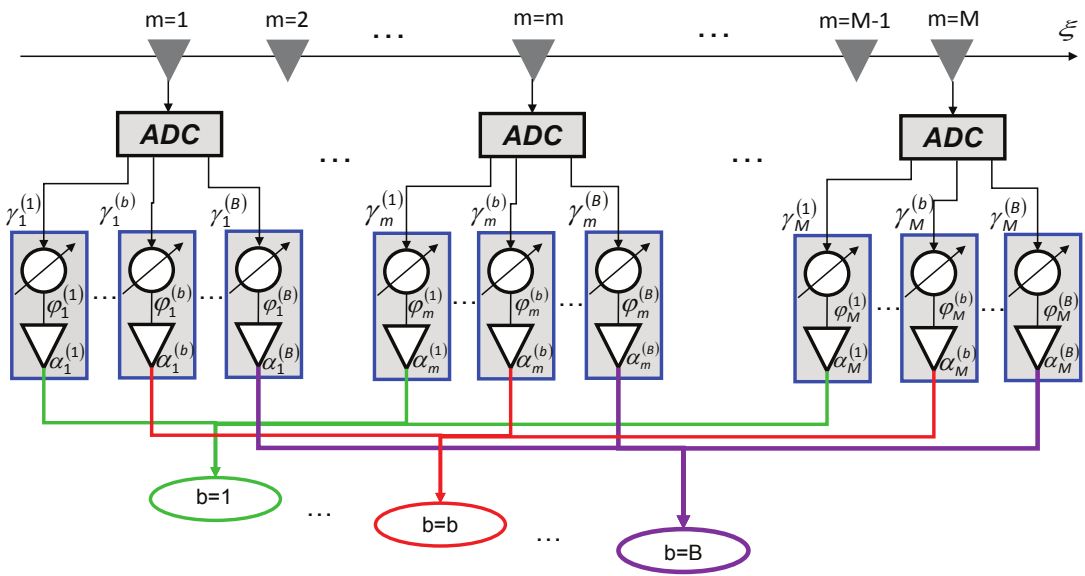

**Fig. 3** - P. Rocca *et al.*, "Conical Multi-Beam Phased Arrays for ..."

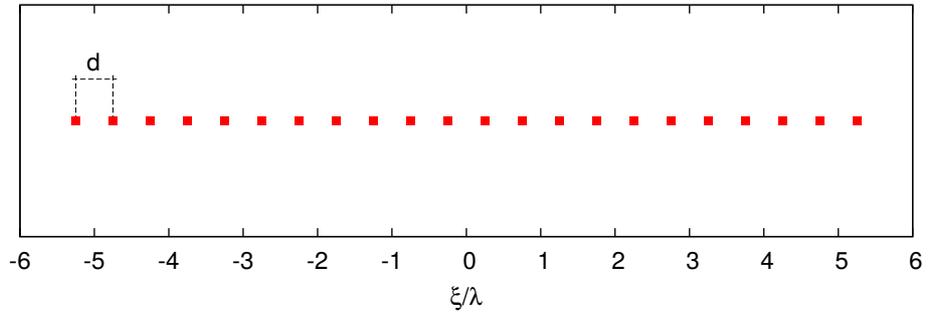

(a)

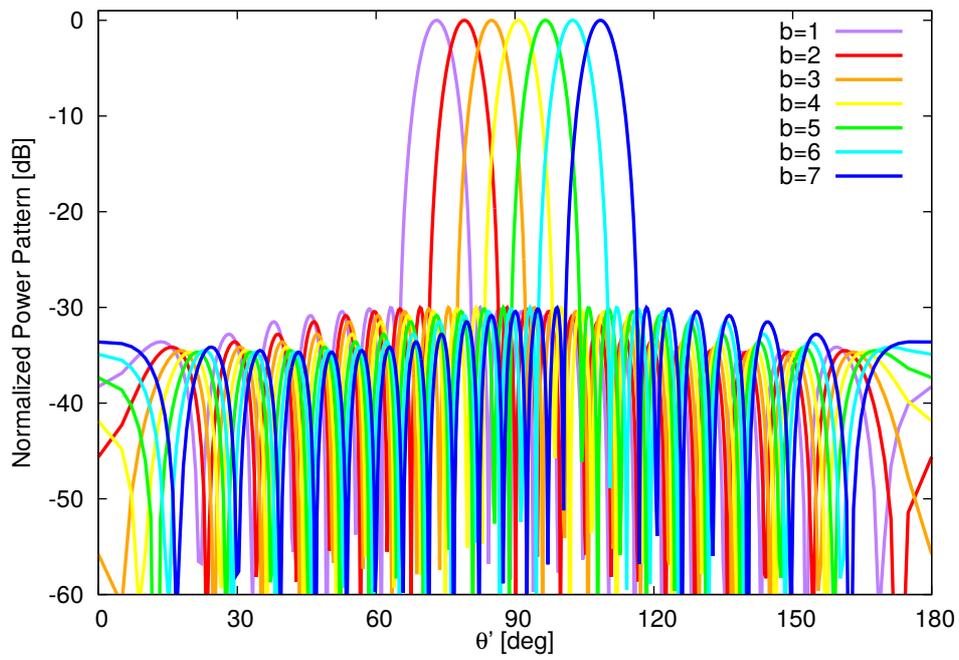

(b)

**Fig. 4 - P. Rocca *et al.*,** "Conical Multi-Beam Phased Arrays for ..."

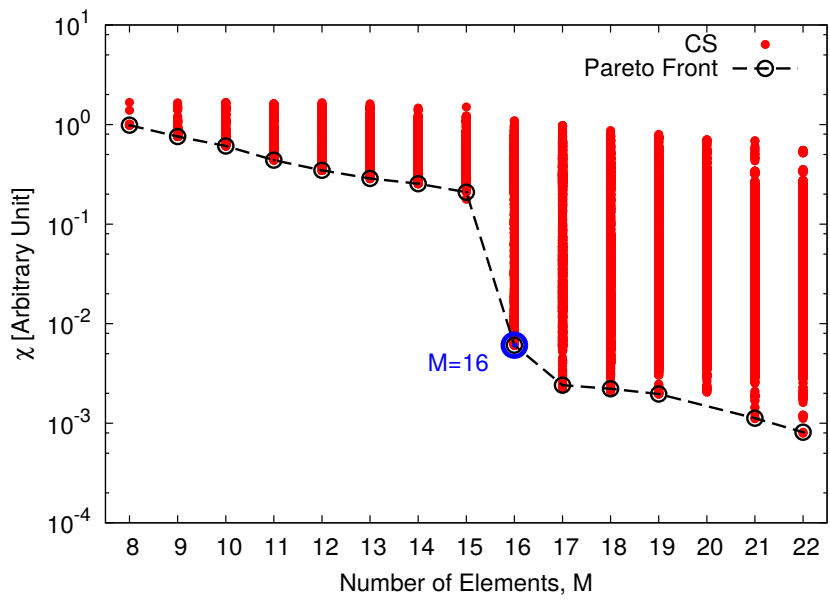

**Fig. 5 - P. Rocca *et al.*,** "Conical Multi-Beam Phased Arrays for ..."

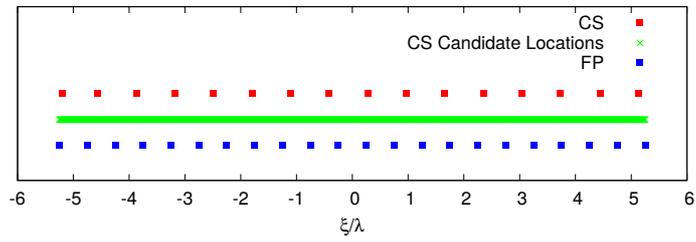

(*a*)

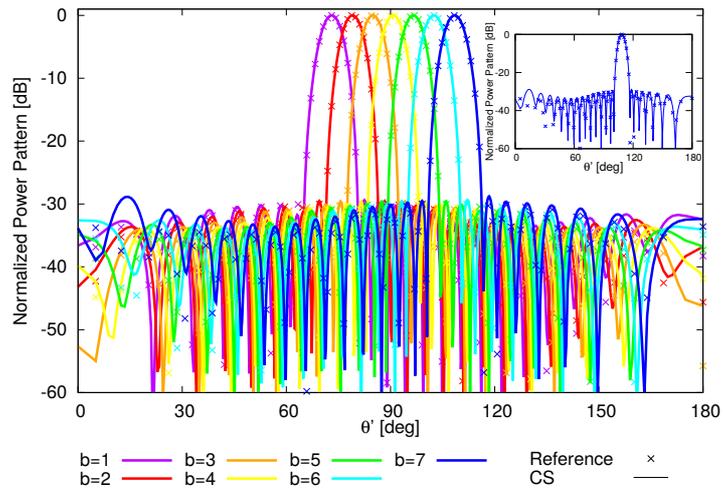

(*b*)

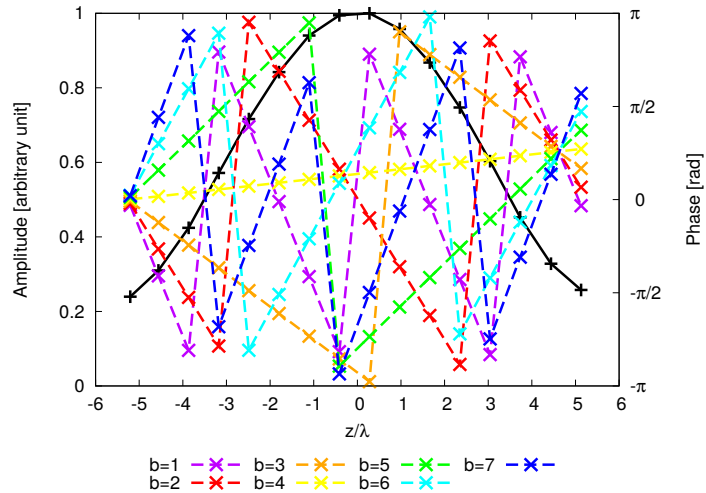

(*c*)

**Fig. 6 - P. Rocca *et al.*,** "Conical Multi-Beam Phased Arrays for ..."

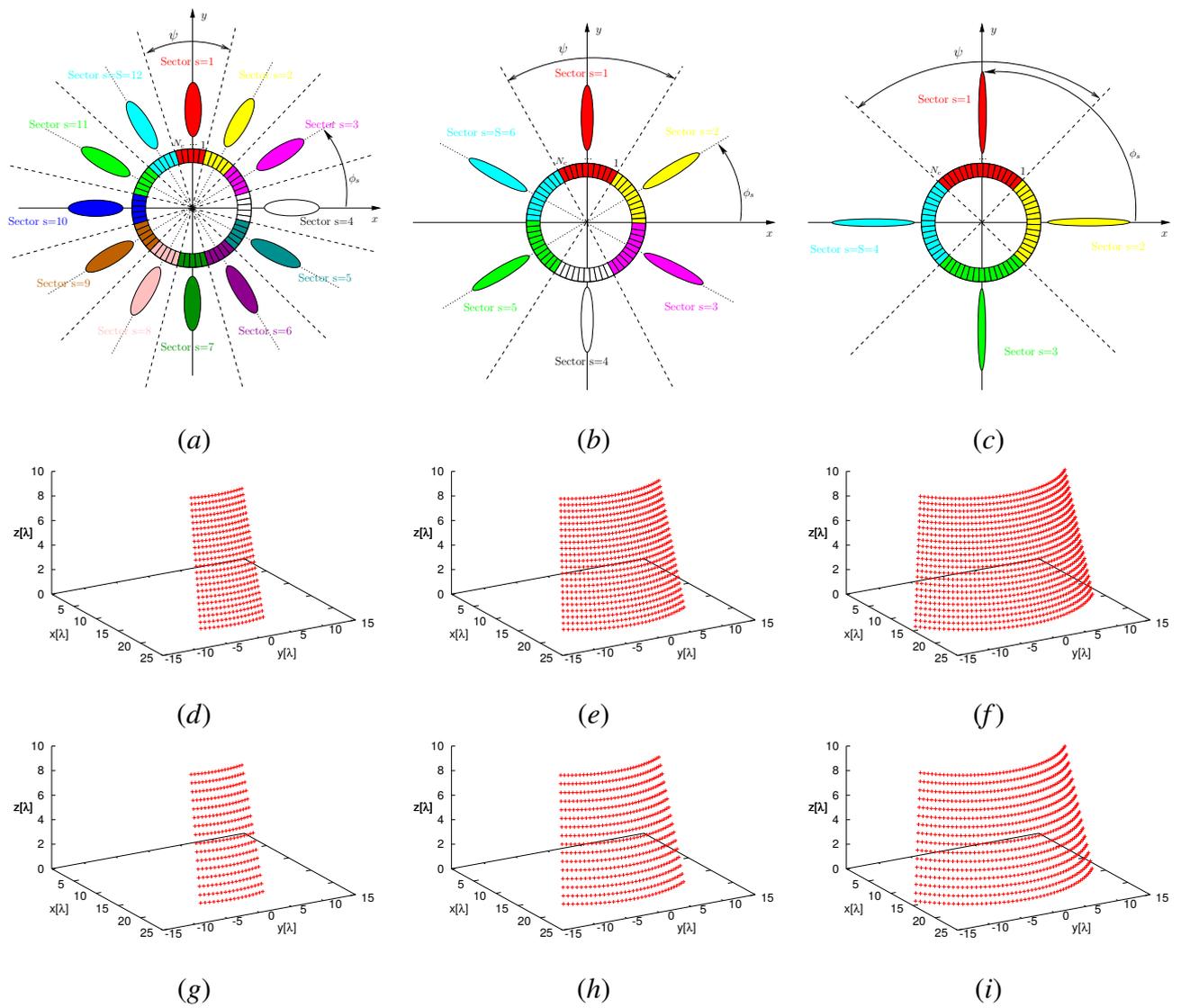

**Fig. 7** - P. Rocca *et al.*, "Conical Multi-Beam Phased Arrays for ..."

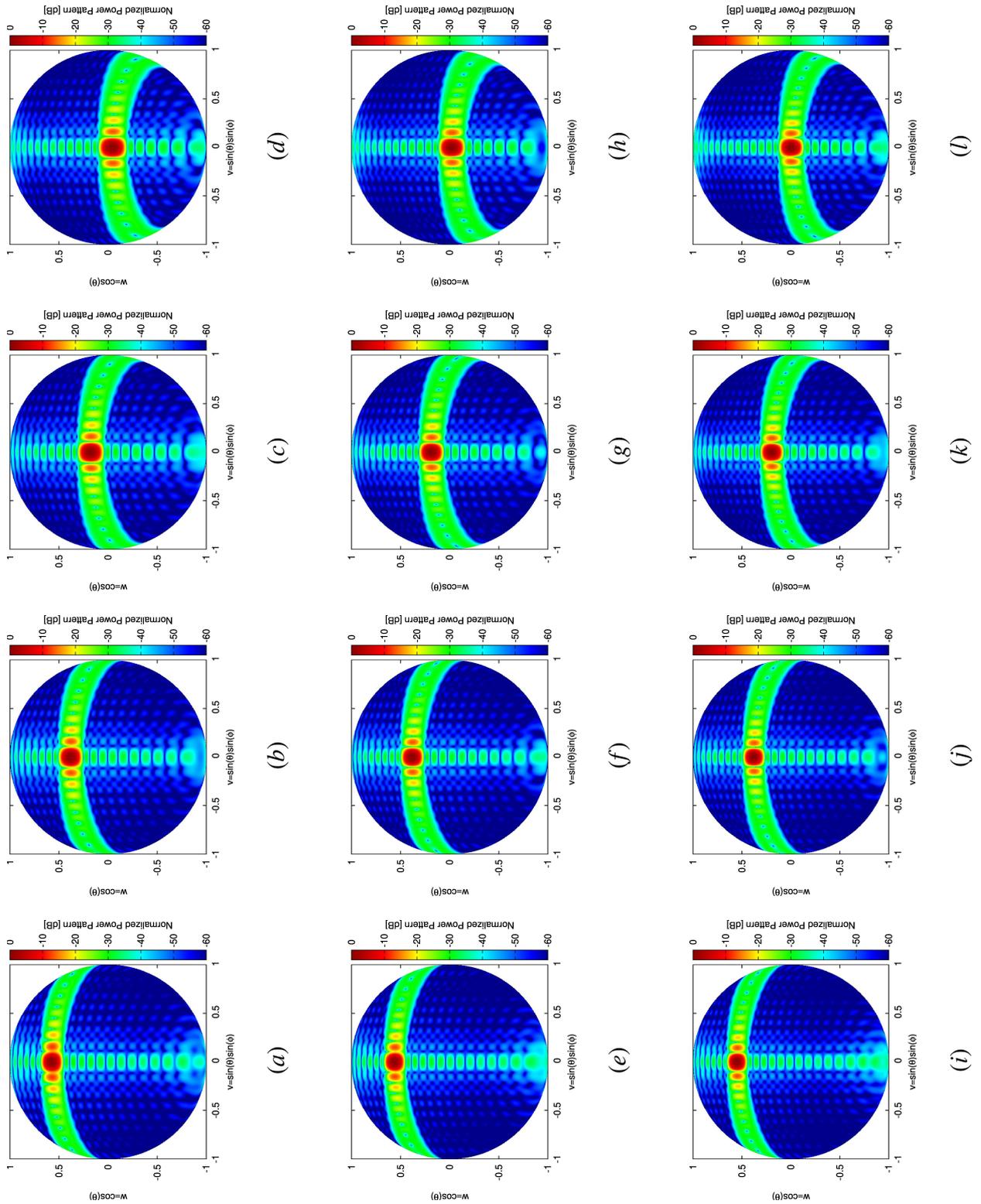

**Fig.** 8 - **P. Rocca** *et al.*, "Conical Multi-Beam Phased Arrays for ..."

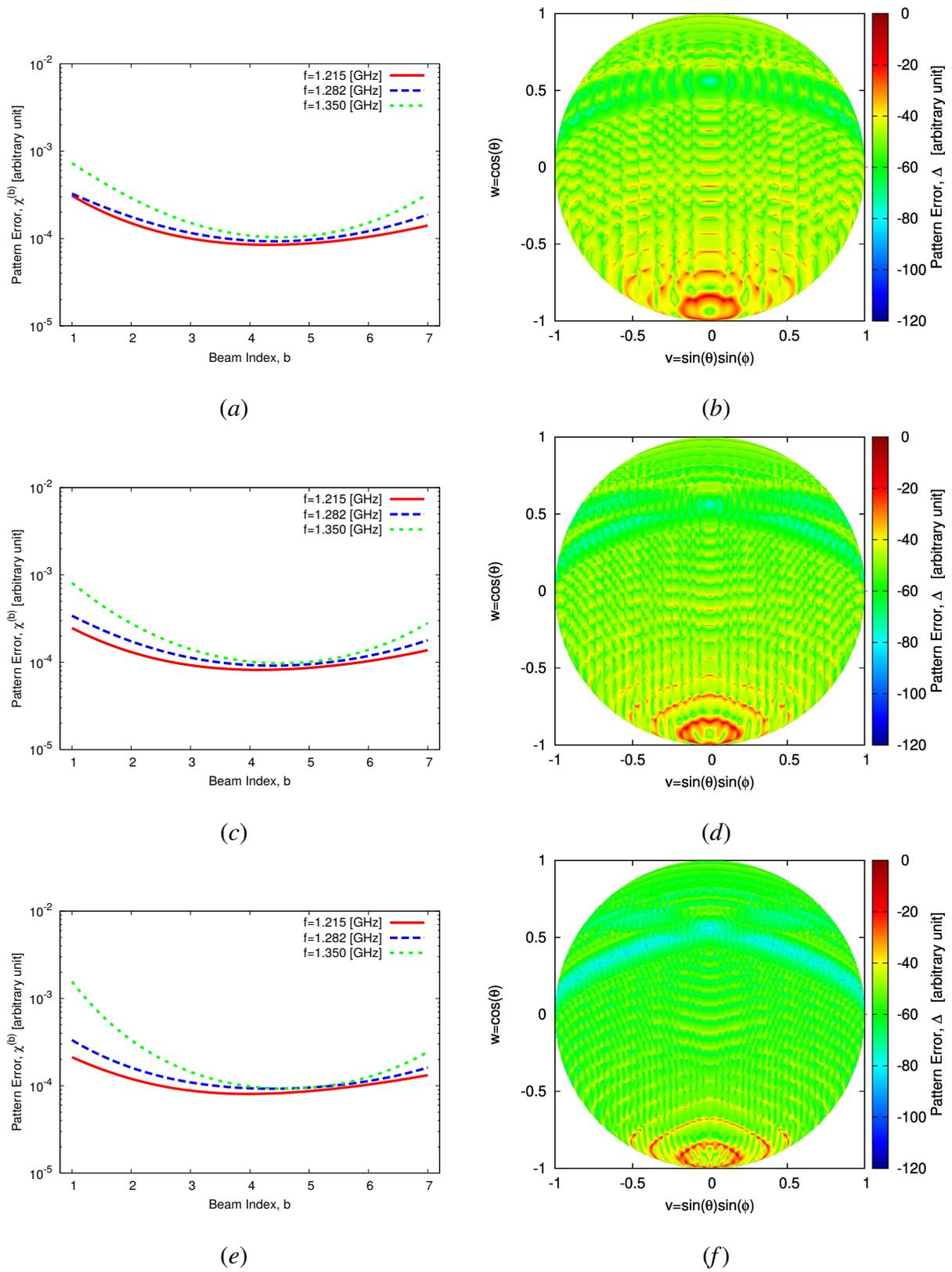

(a) (b)
(c) (d)
(e) (f)

**Fig. 9 -** **P. Rocca *et al.*,** "Conical Multi-Beam Phased Arrays for ..."

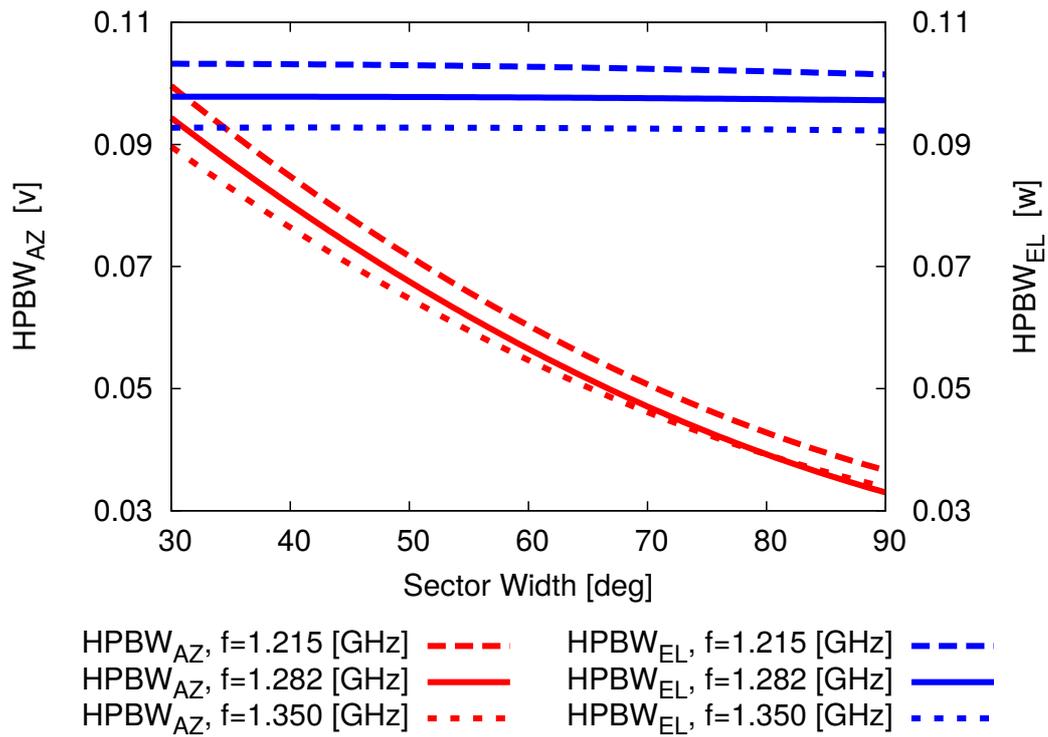

(a)

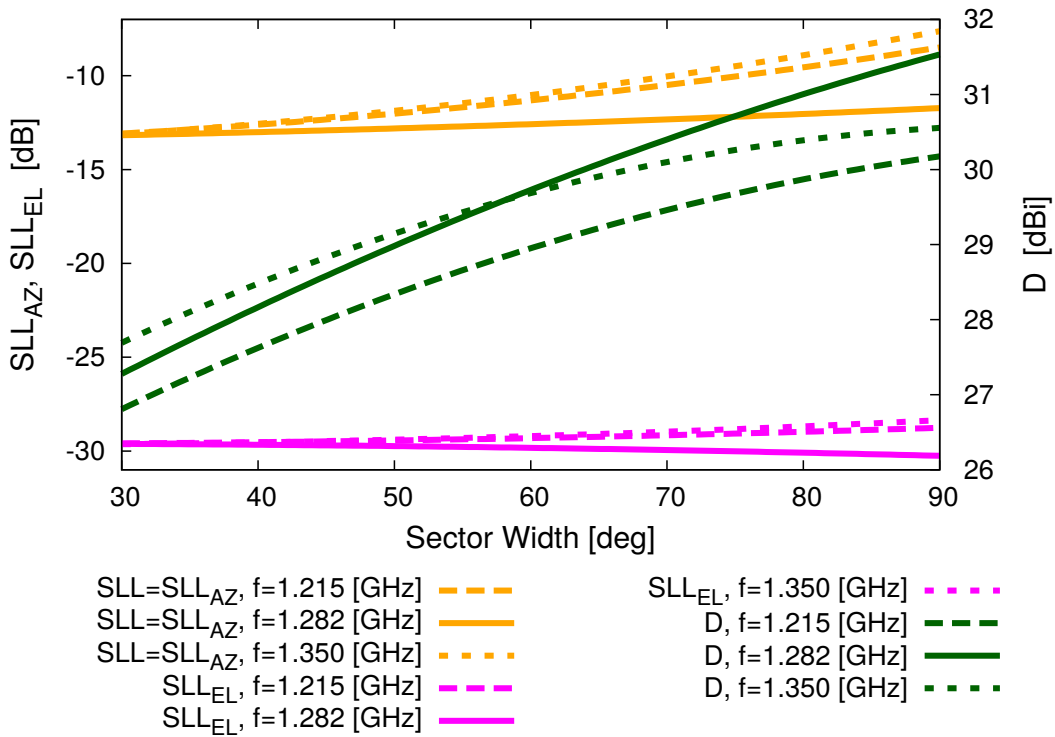

(b)

Fig. 10 - P. Rocca *et al.*, "Conical Multi-Beam Phased Arrays for ..."

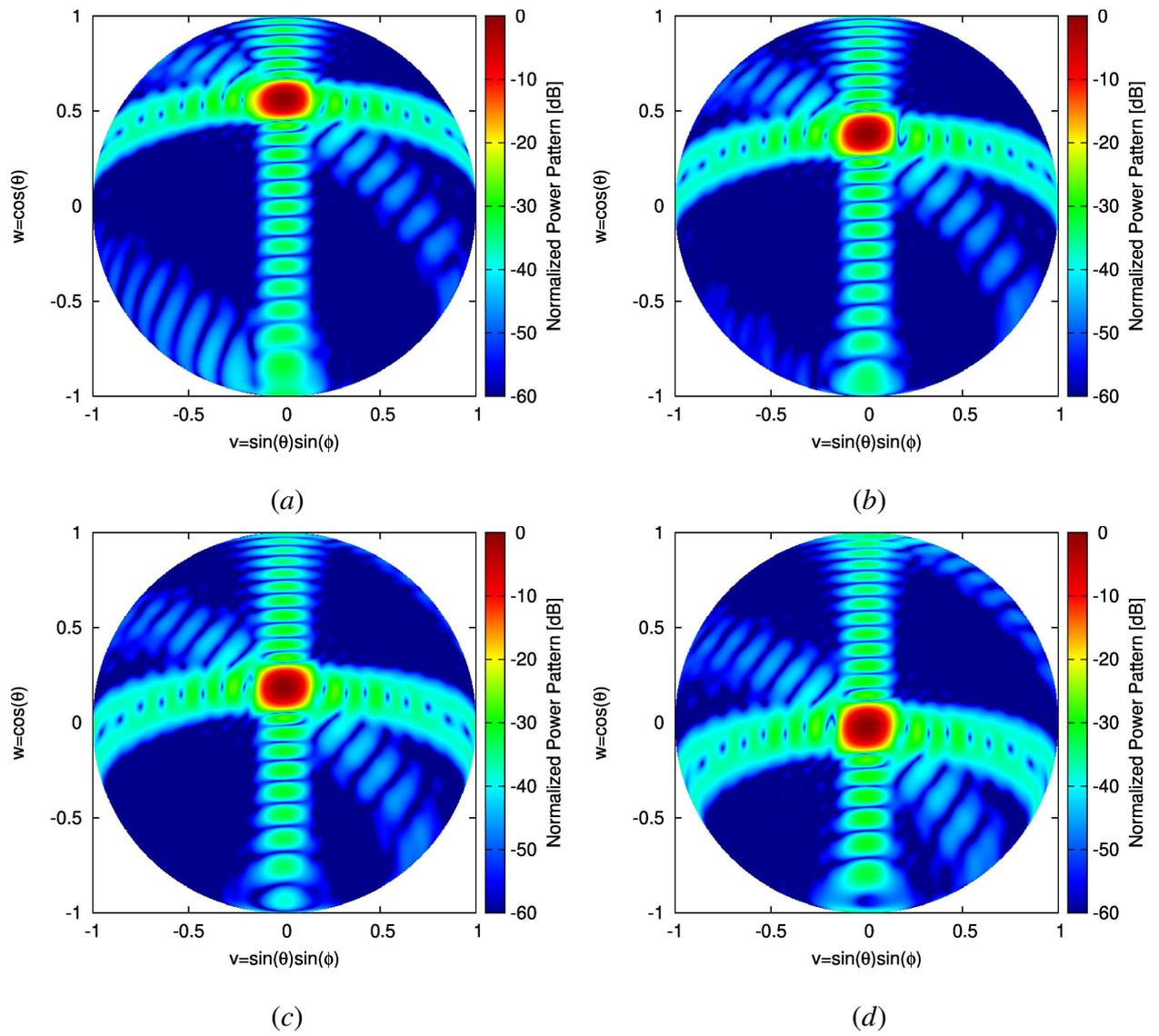

(a)

(b)

(c)

(d)

**Fig. 11** - P. Rocca *et al.*, "Conical Multi-Beam Phased Arrays for ..."

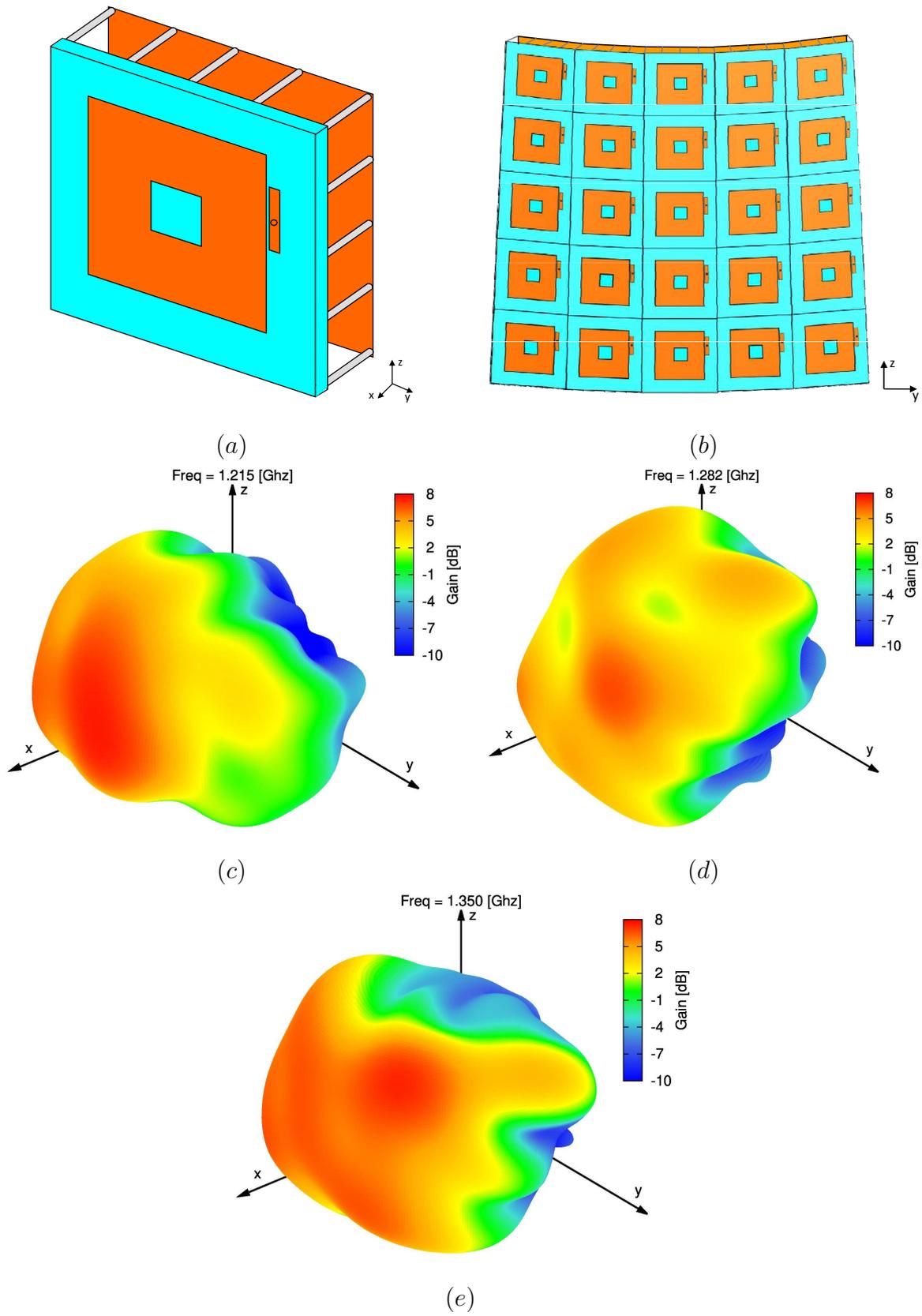

**Fig. 12** - **P. Rocca *et al.*,** "Conical Multi-Beam Phased Arrays for ..."

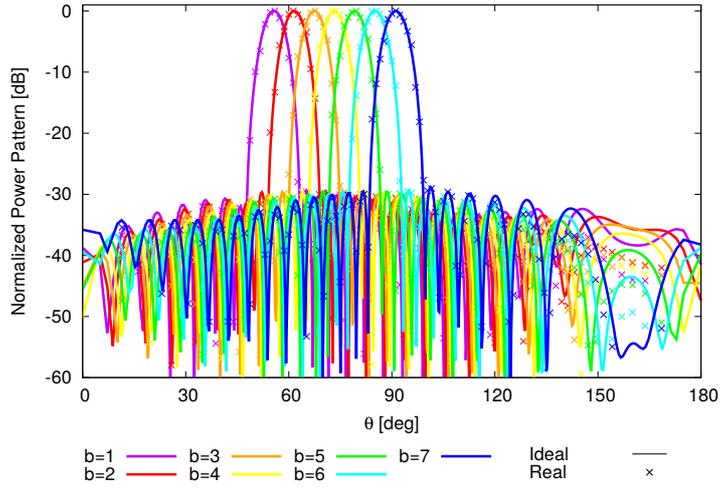

(*a*)

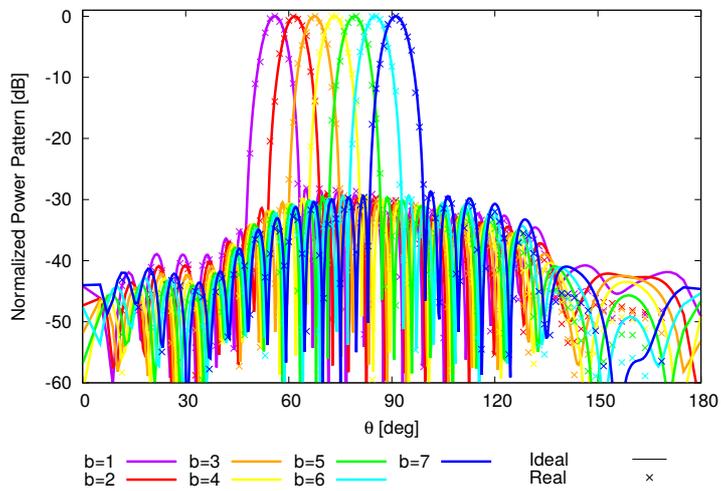

(*b*)

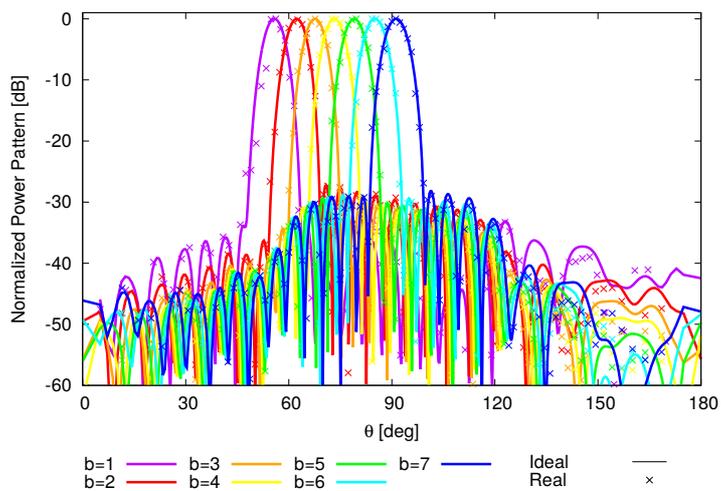

(*c*)

**Fig. 13 - P. Rocca *et al.*,** "Conical Multi-Beam Phased Arrays for ..."

| $b$ | $\theta'^{(b)}$ [deg] | $-3dB_{left}$ [deg] | $-3dB_{right}$ [deg] | $HPBW$ [deg] |
|---|---|---|---|---|
| 1 | 73.07 | 70.00 | 76.09 | 6.08 |
| 2 | 79.06 | 76.09 | 82.00 | 5.92 |
| 3 | 84.93 | 82.00 | 87.85 | 5.83 |
| 4 | 90.75 | 87.85 | 93.65 | 5.81 |
| 5 | 96.57 | 93.65 | 99.50 | 5.85 |
| 6 | 102.46 | 99.50 | 105.45 | 5.95 |
| 7 | 108.47 | 105.45 | 111.55 | 6.13 |

**Tab. I - P. Rocca *et al.*,** "Conical Multi-Beam Phased Arrays for ..."

| $b$ | $SLL$ [dB] | $D$ [dBi] | $HPBW$ [deg] | $\chi^{(b)}$ |
|---|---|---|---|---|
| Reference | | | | |
| | $-30.00$ | 12.77 | 6.08 | |
| CS | | | | |
| 1 | $-29.39$ | 12.73 | 6.12 | $7.90 \times 10^{-3}$ |
| 2 | $-29.51$ | 12.74 | 5.96 | $5.71 \times 10^{-3}$ |
| 3 | $-29.60$ | 12.75 | 5.87 | $4.75 \times 10^{-3}$ |
| 4 | $-29.61$ | 12.75 | 5.85 | $4.61 \times 10^{-3}$ |
| 5 | $-29.53$ | 12.75 | 5.87 | $5.01 \times 10^{-3}$ |
| 6 | $-29.36$ | 12.74 | 5.99 | $5.97 \times 10^{-3}$ |
| 7 | $-28.79$ | 12.77 | 6.13 | $8.58 \times 10^{-3}$ |

**Tab. II - P. Rocca *et al.*,** "Conical Multi-Beam Phased Arrays for ..."